\documentclass[journal]{vgtc}                 

\graphicspath{{figures/}{pictures/}{images/}{./}} 
\usepackage{times}                     
\usepackage{tabu}                      
\usepackage{booktabs}                  
\usepackage{lipsum}                    
\usepackage{mwe}                       
\usepackage{mathptmx}                  
\usepackage{amsmath} 
\onlineid{1574}

\vgtccategory{Research}

\title{DP-LENS: A Density-Aware Polyfocal Lens with Topology-Driven Auto-Routing for Occlusion Management in Immersive 3D Analytics}

\author{%
  \authororcid{Nieyu Cao}{0009-0000-6741-5946},
  \authororcid{Xian Wang}{0000-0003-1023-636X}, and
  \authororcid{Lik-Hang Lee}{0000-0003-1361-1612}
}

\authorfooter{
  \item
  Nieyu Cao is with The Hong Kong Polytechnic University.
  E-mail: nieyucao.cao@connect.polyu.hk.
  \item
  Xian Wang is with The Hong Kong Polytechnic University.
  E-mail: xiann.wang@connect.polyu.hk.
  \item
  Lik-Hang Lee is with The Hong Kong Polytechnic University.
  E-mail: lik-hang.lee@polyu.edu.hk (corresponding author).
}

\teaser{
  \centering
  \includegraphics[width=\linewidth]{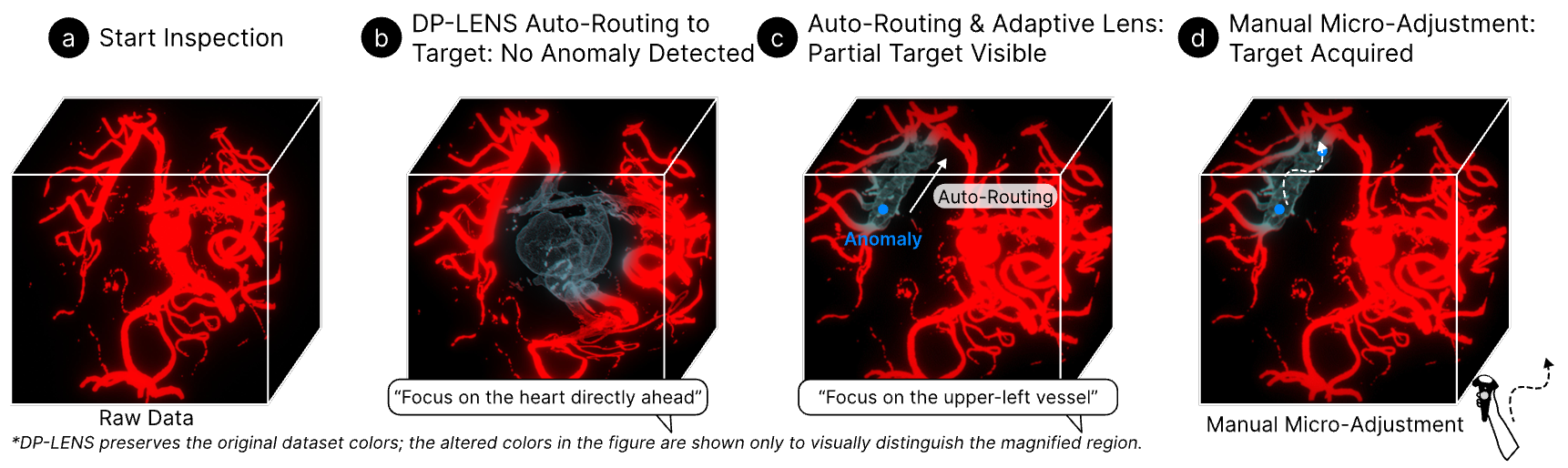}
  \vspace{-6mm}
\caption{The DP-LENS framework synergizes topology-driven auto-routing (initiated via a supplementary LLM-based voice selector) with manual micro-adjustment for exploring heavily occluded 3D datasets. (a) Inspecting a vascular network. (b) The user utilizes a voice command to select a rough target region. (c) The auto-routing algorithm directs the lens to the region, partially revealing a hidden anomaly. (d) The user transitions to manual 6-DOF manipulation for precise micro-adjustment, fully exposing and acquiring the target.}
  \label{fig:teaser}
}

\abstract{
Immersive environments, e.g., virtual reality (VR), offer a unique approach to exploring complex 3D datasets, where data is often heavily occluded and exploration incurs a high cognitive load. We propose DP-LENS, a density-aware polyfocal fisheye lens equipped with topology-driven auto-routing. While preserving peripheral context through geometric deformation and 3D perspective techniques, it enables users to explore 3D data with a lower cognitive load. To facilitate hands-free macro-navigation, we integrate a Large Language Model (LLM) to serve as a supplementary voice-based target selection tool that initiates the auto-routing algorithm. Two user studies with 34 participants investigate the potential benefits of this system. Our first study ($N=18$) compared the manual DP-LENS against two industry-standard baselines (i.e., World-in-Miniature and volumetric slicing) in heavily occluded 3D datasets. The results show that DP-LENS significantly reduced cognitive load, decreased completion time, and improved user preference. The second study ($N=16$) compared the topology-driven auto-routing system (initiated via voice commands) with a fully manual DP-LENS. The results show that the auto-routing system improved task efficiency, further reduced cognitive load, and garnered higher user preference. Furthermore, the auto-routing partially decoupled exploration efficiency from the physical dimensions of the data and mitigated physical fatigue to some extent. Based on the findings, we proposed design implications to inform the development of more spatially scalable and low-fatigue interactions for future 3D visual analytics systems.
} 


\begin{document}

\maketitle

\section{Introduction}
Immersive Analytics (IA) utilizing Head-Mounted Displays (HMDs) is increasingly vital for comprehending complex 3D spatial data. Across diverse domains, such as analyzing cosmological simulations in astronomy~\cite{Cautun2013, Zhao2024}, diagnosing vascular pathologies in medicine \cite{Meyer2008,Preim2013,Oeltze2006}, and tracking particle flows in fluid dynamics~\cite{Laramee2004,Chen2008,Garth2009}. Domain experts have to frequently transition between global overviews and local microscopic details~\cite{Meyer2008,cordeil2019iatk}. However, interacting with such dense, volumetric datasets inherently introduces severe spatial occlusion issues and user workload.

To address this challenge, several studies have introduced lens interaction techniques based on the Focus+Context paradigm. However, these techniques primarily focus on the visual filtering of data with pre-defined labels or multivariate attributes. For example, the Decal-lenses proposed by Rocha et al.~\cite{Rocha2018} are mainly used to manage and switch between pre-computed multivariate data layers (e.g., temperature, pressure) on complex 3D surfaces. Similarly, the 3De lens designed by Mota et al.~\cite{Mota2018} focuses on visual switching between different, pre-generated geometric representations (such as surfaces and streamlines). These approaches heavily rely on existing semantic presets or pre-computed layers of the data. In contrast, in many real-world exploratory tasks, data is completely raw and unlabeled. How to seamlessly and dynamically balance focal observation (Focus) and spatial context (Context) for such raw volumetric data, particularly relying on physical deformation rather than semantic filtering, remains a critical, underexplored challenge. 

The explorations of dense volumetric data in immersive VR environments pose two major challenges. \textbf{(1) How to maintain spatial context while accessing microscopic details.} In the dense world, users may experience difficulty in mentally reconstructing the global structure due to the loss of topological connectivity, resulting from either the context being removed by slicing or the focus being decoupled in a separate view~\cite{Elmqvist2008,Stoakley1995,Zhao2022LWiM}. In 3D environments that lack inherent spatial references, this problem becomes even more pronounced due to the \textit{Desert Fog} effect, where local details lack global grounding~\cite{Julier2000,gruenefeld2019comparing}. Compounding this cognitive disorientation is the physical challenge of traversing these expansive, dense datasets. 
\textbf{(2) How the user can accurately and efficiently navigate to distant targets.} The user and the Region of Interest (ROI) may be separated by a large scale in a room-scale environment, and occlusion between them could deteriorate their interaction, driven by dense particle clouds or the sheer volume of data~\cite{Bermejo2021Button}. Therefore, the user may not be able to directly reach the desired target without experiencing significant fatigue.

The integration of Focus+Context paradigms serves as a prominent solution to the aforementioned challenges in exploring volumetric data. Employing a distortion lens that seamlessly blends focal details aids in preserving the continuity and topology of the data. When users can perceive the surrounding context while focusing, it improves orientation and comprehension of the structure, enabling more effective exploration~\cite{Cockburn2009}. Studies have shown that integrating focus and context significantly improves the efficiency of exploration tasks by providing continuous visual information~\cite{Carpendale1997,tominski2017interactive}. Supporting density-aware mechanisms in a single user task also allows users to identify and localize target clusters out of view more accurately in VR and AR~\cite{Yu2012,Zhao2024,Zhao2026ScaleFree}. However, these studies do not explore the ubiquitous yet fundamental task of density-driven interaction in the 3D lens, and most proposed techniques strictly focus on manual geometric deformation. Moreover, there is limited discussion of existing techniques operating in room-scale environments. To the best of our knowledge, prior research has rarely discussed scenarios where lens operations are assisted by \textbf{voice-driven LLM agents} to automate navigation and reduce physical load.

To address these challenges, we propose a unified framework, \textit{Density-Aware Polyfocal Fisheye (DP-LENS)}, that combines manual dexterity with intelligent automation for dense 3D data exploration and enables context-preserving exploration of occluded internal structures. Supported by a unified point-based data architecture, DP-LENS broadly leverages underlying density fields to execute intelligent magnetic snapping and morphological deformations across diverse topologies. Furthermore, recognizing that continuous 6-DOF manual manipulation in room-scale, feature-sparse environments inherently induces severe physical fatigue and cognitive overhead, we augment DP-LENS with a \textit{topology-driven auto-routing} system~\cite{zhang2023complete,zhang2023survey,Yang2024Prompt}. By utilizing a Large Language Model (LLM) as a supplementary voice-based target selector, this system translates natural language intents into target regions, which subsequently initiates our automated long-distance trajectory planning. Ultimately, this augmentation alleviates the burden of continuous spatial piloting, allowing users to allocate more of their cognitive resources toward high-level data analysis.

The primary contributions of this paper are threefold: \textbf{(1) DP-LENS Technique:} A novel Density-Aware Polyfocal Fisheye interaction that integrates volumetric probes with adaptive distortion. Unlike destructive clipping, our technique is designed to leverage underlying density fields to resolve severe 3D occlusion while strictly preserving the topological connectivity of complex datasets. \textbf{(2) Topology-Driven Auto-Routing with Voice Selection:} An automated navigation framework that bridges a supplementary LLM-based voice selector with pre-computed spatial density analysis. By translating abstract natural language intents into target regions, the system performs real-time density querying to execute automated, collision-free lens trajectories, effectively mitigating the impact of physical dimensions on exploration efficiency within standard room-scale boundaries. \textbf{(3) Empirical Validation and Open-Source Toolkit:} Comprehensive evidence from two controlled experiments demonstrating the efficacy of DP-LENS and its auto-routing in reducing cognitive load and physical fatigue. To support open science and provide a reusable foundation for the community, we contribute a reproducible software toolkit, comprising our high-performance GPU compute shaders, the multimodal LLM orchestration logic, and the standardized 3D evaluation datasets.

\section{Related Work}

\paragraph{Navigation and Exploration in Abstract 3D Environments}


Modern immersive analytics enables spatial exploration of 3D data in VR and AR by leveraging stereoscopic depth cues for tasks such as point cloud clustering~\cite{Kraus2022,Zhao2024}, vortex tracing in fluid dynamics~\cite{Laramee2004,Chen2008,Garth2009}, and vascular blockage localization~\cite{Meyer2008,Preim2013,Oeltze2006}. Classical navigation techniques, including zooming and Overview+Detail methods such as World-in-Miniature (WIM)~\cite{Cockburn2009}, support access to local detail but often compromise awareness of global structure. More recent approaches, such as collaborative WIM systems~\cite{Zhao2022LWiM,Wang2025TeamPortal} and adaptive layout techniques~\cite{Li2023ImmerView,bauer2021multi}, improve global context, yet they frequently require repeated gaze shifts between overview and detail, which can interrupt analytical flow. To better balance local detail and global context, prior work has explored Focus+Context techniques. Early methods based on global 3D distortion~\cite{Carpendale1997} sought to unify detailed and contextual views within a single representation, but introduced substantial spatial deformation. Interactive lens approaches~\cite{tominski2017interactive,o2024immersive} reduced this problem by confining transformations to localized regions; however, they remain susceptible to occlusion and can disrupt visual continuity.

These limitations become more pronounced in fully 3D mid-air environments, where lens placement and manipulation introduce additional ergonomic and interaction challenges. Unconstrained 6-DOF manual interaction often lacks physical support, leading to a phenomenon widely recognized in HCI as the ``Gorilla Arm'' effect. Specifically in the context of spatial visualization, Besançon et al.~\cite{besanccon2021state} highlighted that unsupported mid-air gestures are prone to natural hand jitter and decreased precision, while Yu et al.~\cite{yu2010fi3d} noted that prolonged spatial manipulation quickly induces physical fatigue. 
To address these issues, recent research has proposed new interactive technologies, including physical tablets and tangible volumes~\cite{Langner2021}, haptic guidance~\cite{issartel2016tangible,Zhao2024SpatialTouch}, visual 3D gizmos~\cite{taibo2024immersive}, and multimodal interaction strategies~\cite{Yuan2023MEinVR,Zhang2020Hand}. However, most existing approaches still rely on explicit manual control, physical props, or visual feedback, and only limited attention has been given to reducing fatigue and spatial misalignment during prolonged unsupported interaction. DP-LENS addresses this gap by supporting navigation and inspection in abstract, complex 3D environments while reducing the burden of manual operation.

\paragraph{Occlusion Management and Context-Aware Interaction}
Occlusion management is central to the exploration of dense 3D data. Elmqvist and Tsigas classified occlusion management techniques into multiple viewports, tour Planners, projection distorters, virtual X-ray, and volumetric Probes~\cite{Elmqvist2008}. For raw volumetric data without predefined semantic labels, the most relevant paradigms are multiple viewports, virtual X-ray, and volumetric probes. WIM and related multi-view point techniques separate overview from detail and can provide global context~\cite{Cockburn2009,Zhao2022LWiM,Wang2025TeamPortal}, but they require attention shifts and become visually cluttered for dense structures. In contrast, virtual X-ray approaches based on geometric slicing or clipping expose internal regions in place~\cite{Li2023}, yet they do so by removing foreground structures and thereby disrupting spatial continuity. This trade-off motivates context-aware alternatives that preserve both visibility and structural coherence. Prior work has shown that density-aware interaction can support more effective selection in dense 3D environments, as demonstrated by CloudLasso and MeTACAST~\cite{Yu2012,Zhao2024}. These results suggest that the density field can serve as an implicit cue for interaction and target inference. Building on this insight, our work treats density awareness not as an auxiliary selection aid, but as the basis of a volumetric probe for non-destructive occlusion management. In this way, DP-LENS combines the localized access of volumetric probes with context-preserving reveal, aiming to improve internal visibility without dividing attention or sacrificing topological connectivity~\cite{zimmermann2025multi}.

\paragraph{LLM-based Navigation Assistance}

Recent research in embodied navigation has explored Large Language Models (LLMs) as a high-level reasoning layer for translating natural-language instructions into executable navigation plans~\cite{Ahn2022,Zhou2024,Song2025}. 
Prior work has shown that LLM-based agents can support instruction decomposition~\cite{Ahn2022}, skill orchestration~\cite{Song2023}, and stepwise decision-making~\cite{Zhou2024}, improving adaptability in previously unseen tasks.
More recent approaches further enhance navigation, for example, Song et al.~\cite{Song2025} and Fan et al.~\cite{Fan2025} conducted studies on enhancing spatial awareness through multi-granularity dynamic memory and topological scene maps~\cite{liu2025dgdiff}. 
To further bridge the gap between discrete and continuous settings, recent methodologies have expanded into continuous environments and global context understanding, including spatial-temporal Chain-of-Thought (CoT) reasoning~\cite{Qiao2025}. However, these methods are primarily designed for semantically structured physical environments, where navigation targets are defined by explicit landmarks, objects, or rooms. In immersive analytics, the navigation problem is fundamentally different. Analysts explore abstract, dense volumetric spaces that often lack discrete semantic anchors, requiring interpretation of spatial density patterns and smooth 6-DoF movement during inspection. In this setting, continuous manual navigation can introduce substantial physical and cognitive burden. Our work adopts LLMs not as a general embodied agent, but as a navigation assistant that interprets user intent and automates long-range lens movement in dense 3D data. This design complements DP-LENS by delegating macro-navigation to language-guided planning while preserving manual control for local, precise inspection.

\section{DP-LENS: Design and Implementation}

To address the physical and cognitive bottlenecks of exploring densely occluded 3D environments, our architecture synergizes a density-aware volumetric rendering engine with a topology-driven auto-routing mechanism, supported by a supplementary voice selector for intent parsing.

\subsection{Density-Aware Volumetric Probe (DP-LENS)}
\label{sec:probe_design}

Unlike traditional planar clipping, DP-LENS is designed as an intelligent volumetric probe that dynamically displaces occluding structures while preserving topological context.

\paragraph{Topological Skeletonization and Flow Repair} To grant the system awareness of the underlying data structure, we first compute a continuous 3D density field $\rho(\mathbf{r})$ from the discrete raw points using Kernel Density Estimation (KDE)~\cite{Zhao2026ScaleFree}. To ensure real-time interaction, this computation is highly parallelized using GPU Compute Shaders. Based on the scalar field $\rho(\mathbf{r})$, we extract the topological skeleton (centerlines) to guide the lens deformation. However, in complex or noisy datasets (e.g., fractured vascular scans), extracted skeletons often suffer from disjointed segments. To ensure continuous navigation, we propose a Gradient-Driven Flow Repair algorithm. Let the current state of a virtual probe be its position $\mathbf{p}$ and tangent direction $\mathbf{d}$. The probe extends outwards by evaluating a heuristic energy function $E(\mathbf{v})$ to find the optimal candidate connecting segment, defined as $E(\mathbf{v}) = \frac{\alpha}{\|\mathbf{p} - \mathbf{v}\|} + \beta \left( 1 - \frac{\theta(\mathbf{d}, \mathbf{d}_v)}{\pi} \right)$, where $\mathbf{v}$ is the starting node of the candidate segment, $\mathbf{d}_v$ is its tangent, and $\theta$ calculates the angular difference. $\alpha$ and $\beta$ are weighting coefficients for distance and angular momentum. The probe ``swims'' along the direction of the local density gradient $\nabla \rho(\mathbf{p})$, effectively bridging topological gaps to generate a fully connected structural centerline.

\paragraph{Continuous Polyfocal Deformation and Auto-Scaling} Traditional geometric magnifiers (e.g., spherical blobs) distort space around a rigid single focal point, which significantly limits their versatility, particularly when tracking complex or elongated structures. To provide greater adaptability and generality across diverse data topologies, DP-LENS dynamically generates a continuous polyfocal visual tube $T$ along the previously extracted topological skeleton $L(\mathbf{r})$. By dynamically querying the pre-computed density field $\rho(\mathbf{r})$ to guide the focal center toward this skeleton, the lens intrinsically yields a ``magnetic snapping'' effect, significantly reducing the positional manual precision required by the user. For the visual deformation, we extend the classic Sarkar-Brown fisheye magnification model~\cite{sarkar1994graphical}. For a given point $\mathbf{x}$, let $\mathbf{x}_{proj}$ be its closest projection on the focal tube $T$, and $r = \|\mathbf{x} - \mathbf{x}_{proj}\|$ be the perpendicular distance. We define the normalized distance as $\tilde{r} = r / R_{safe}$. The displaced radius $r_{new}$ is then modulated by a distortion factor $d$, such that:
\begin{equation}
r_{new} = R_{safe} \frac{(d + 1)\tilde{r}}{d\tilde{r} + 1}
\end{equation}
Crucially, the maximum expansion radius $R_{safe}$ is dynamically governed by a Density-Aware Auto-Scaling mechanism. As the focal tube translates, the system casts discrete detection rays orthogonal to the tangent of the centerline. The safe radius at any position $\mathbf{c}$ on the tube is determined by the minimum distance to the topological boundary where the scalar density $\rho(\mathbf{r})$ drops below a user-defined iso-threshold $\tau$, calculated as:
\begin{equation}
R_{safe}(\mathbf{c}) = \min_{\mathbf{u}} \{ s \mid \rho(\mathbf{c} + s\mathbf{u}) < \tau, \, \mathbf{u} \cdot \mathbf{t} = 0 \}
\end{equation}
where $\mathbf{u}$ represents orthogonal sampling directions, $s$ is the scalar distance along the ray, and $\mathbf{t}$ is the local tangent vector. This formulation intelligently halts the lens expansion upon detecting structural walls, thereby avoiding severe topological conflicts.

\paragraph{Context-Preserving X-Ray Rendering} To reveal internal structures without permanently obliterating the foreground, DP-LENS integrates a view-dependent contextual suppression pipeline. Unlike standard binary volumetric clipping~\cite{Weiskopf2003,Preim2013} (evaluated later as our V-SLICE baseline), which relies on an absolute boolean function $\mathrm{discard}(\mathbf{x})$, our approach calculates a continuous occlusion penalty. Let $\mathbf{C}$ be the camera position and $\mathbf{v} = \frac{\mathbf{x} - \mathbf{C}}{\|\mathbf{x} - \mathbf{C}\|}$ be the view ray casting towards a particle $\mathbf{x}$. We construct a visual frustum bounded by the magnified focal tube. The orthogonal distance $D$ from the tube's central axis to the view ray is evaluated. The final alpha transparency $\alpha$ of any particle located physically outside the focal tube but within the visual frustum is dynamically attenuated as:
\begin{equation}
\alpha = \alpha_0 \left( 1 - \gamma \cdot \mathcal{S}(R, 0.8R, D) \right)
\end{equation}
where $\alpha_0$ is the base opacity, $\gamma \in [0, 1]$ is the context suppression power, $R$ is the local magnified radius of the tube, and $\mathcal{S}$ represents a smoothstep interpolation function. This alpha-blending strategy ensures occluding structures transition into a soft ``X-Ray/Ghosting'' state rather than disappearing completely, thus preserving the global spatial reference anchored in the user's peripheral vision.

\begin{figure}[tb]
 \centering 
 \includegraphics[width=\columnwidth, alt={A block diagram illustrating the Hybrid Voice-Initiated Auto-Routing pipeline, including voice input processing, scene graph generation, LLM reasoning, and density-aware routing.}]{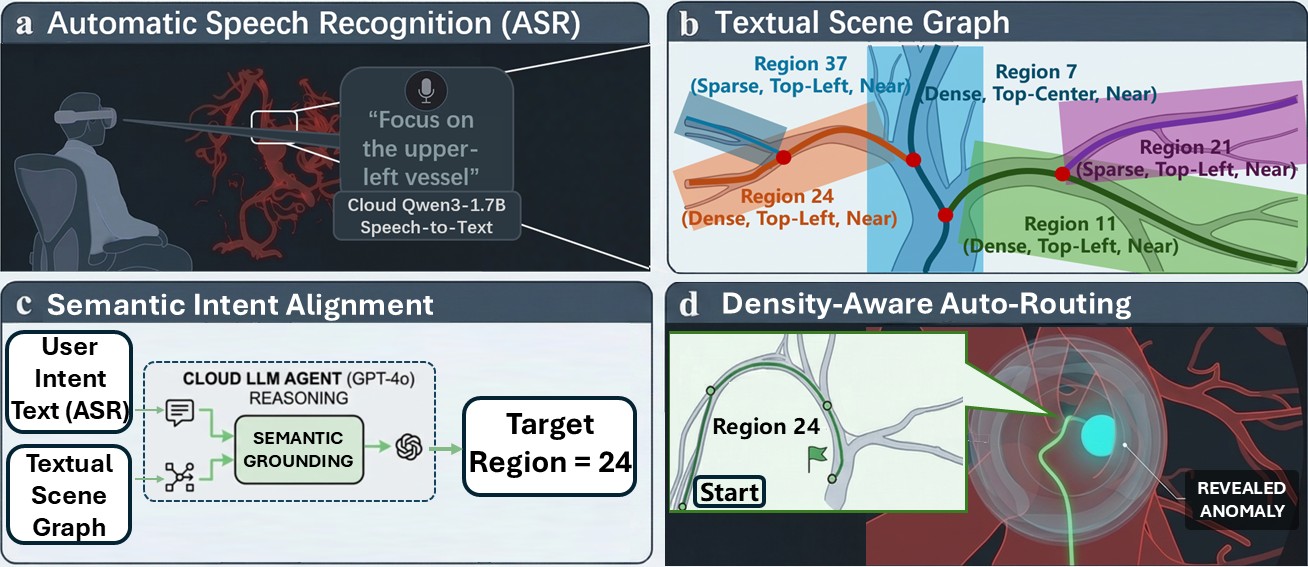} 
\caption{The Hybrid Voice-Initiated Auto-Routing pipeline. (a) ASR transcribes voice input. (b) 3D data is discretized into a Textual Scene Graph. (c) A cloud LLM performs semantic grounding to resolve the target. (d) Density-aware auto-routing guides the lens along the topological skeleton to reveal the anomaly.}
 \vspace{-5mm}
 \label{fig:ai_pipeline}
\end{figure}

\subsection{Hybrid Voice-Initiated Auto-Routing}
\label{sec:ai_nav}

To mitigate physical fatigue during long-distance traversal, we introduce an auto-routing mechanism initiated by an LLM-based voice selector. Although contemporary LLMs excel at semantic reasoning, directly parsing dense, continuous 3D coordinate arrays remains computationally inefficient and prone to spatial hallucination for current state-of-the-art conversational agents. To bridge the modality gap between the continuous physical space and the LLM's discrete text modality, we developed a \textit{Dynamic Spatial Context Injection} pipeline (\cref{fig:ai_pipeline}). 
First, the system performs 3D scene discretization (\cref{fig:ai_pipeline}b). Since the LLM cannot process raw floats directly, the system continuously evaluates the world coordinates $\mathbf{P}_w$ of the extracted vascular segments and projects them into the user's normalized viewport space $\mathbf{P}_{vp} = (x, y, z)$. These spatial coordinates, along with density features, are translated into a lightweight, textual scene graph. Specifically, the positions are discretized into topological semantic tags $S_{tags}(x,y)$ (e.g., yielding ``Top Left'' if $x < 0.4 \land y > 0.6$, ``Center'' if $0.4 \le x \le 0.6 \land 0.4 \le y \le 0.6$, etc.), coupled with depth heuristics and relative density classifications. Specifically, a region is tagged as ``Near'' if its normalized depth falls within the front half of the bounding volume ($\mathbf{P}_{vp}.z < 0.5$), and ``Far'' otherwise. Similarly, to robustly categorize structural density across diverse datasets, the density semantic tag $S_{\rho}$ for a local region at $\mathbf{r}$ is assigned as:
\begin{equation}
S_{\rho}(\mathbf{r}) = 
\begin{cases} 
\text{``Sparse''}, & \text{if } \rho(\mathbf{r}) < 0.5 \bar{\rho} \\ 
\text{``Dense''}, & \text{otherwise} 
\end{cases}
\end{equation}
where $\bar{\rho}$ is the global mean density of the dataset.

Following this spatial-to-text translation, the system executes prompt construction and intent alignment. Upon receiving a voice command, the audio is first transcribed into text via a commercial Automatic Speech Recognition (ASR) API powered by the Qwen3-1.7B model~\cite{Qwen2024Audio} (\cref{fig:ai_pipeline}a). Subsequently, the local VR client constructs a structured system prompt combining the discretized scene tags (e.g., ``\textit{Region 12 is Dense, Top-Left, and Near}'') with the user's transcribed query. The integrated cloud-based LLM (i.e., GPT-4o~\cite{OpenAI2024GPT4o}) then executes multimodal reasoning based on this injected context (\cref{fig:ai_pipeline}c). It aligns the user's fuzzy verbal intent (e.g., \textit{``Focus on the vessel on the top left''}) with the provided scene descriptions, successfully resolving the target (e.g., Region 24) even if the user omits specific density or depth attributes. Finally, the system automatically interpolates the focal center $\mathbf{C}(t)$ along the pre-computed optimal density path $P(t)$ of the selected segment, executing an automated, collision-free lens trajectory (\cref{fig:ai_pipeline}d).

\subsection{System Implementation}
\label{sec:implementation}

The system was developed in Unity (v.2020.3 LTS) using the Meta Interaction SDK. To maintain the strict 90 FPS threshold required for VR comfort, the system employs an optimized GPU-driven pipeline. By loading massive datasets (up to 2 million points) directly into VRAM via ComputeBuffers, we bypass CPU-to-GPU memory transfer bottlenecks, allowing custom geometry shaders to dynamically evaluate context-preserving X-ray fields in real-time. ASR and LLM services are integrated via RESTful HTTP requests. To prevent interaction latency and frame drops, these network requests and multimodal intent parsing execute asynchronously, ensuring continuous rendering and smooth 6-DOF manual manipulation. The system runs on a high-performance workstation (NVIDIA RTX 4070, Intel Core i9) tethered to a Meta Quest 3 headset, utilizing standard 6-DOF controllers for precise ray-casting.To ensure proper depth perception, which is critical for spatial understanding, the pipeline extracts structural normals during the data conversion phase. These normals interact with Unity's directional lighting to provide essential shading and depth cues.

System Scalability and Frame-Time Breakdown. 
To discuss the system's scalability and understand its performance under varying data scales, we measured the frame rates and frame times across different point counts (up to 1.5M points). As summarized in Table \ref{tab:scalability}, at the baseline of 200,000 points, the system averages 118.97$\pm$3.94 FPS (frame time: 8.42 ms) with the lens disabled, and 114.14$\pm$10.29 FPS (8.85 ms) when the lens is fully activated. Furthermore, a stress test on a dataset of 1.5 million points yields 118.90$\pm$4.54 FPS (8.43 ms) without the lens, and 110.68$\pm$15.03 FPS (9.25 ms) with the lens activated. Overall, the dynamic lens deformation and context-preserving rendering introduce an overhead of less than 1 ms per frame. This indicates that the GPU-driven pipeline can gracefully maintain frame rates above the 90 FPS VR comfort threshold even when scaling up to 1.5 million points.

\begin{table}[tb]
\centering
\caption{Scalability and frame-time breakdown (FT = frame time). FPS values are reported as Mean $\pm$ Standard Deviation (SD).}
\label{tab:scalability}
\resizebox{\columnwidth}{!}{%
\begin{tabular}{lcccc}
\toprule
\textbf{Dataset Size} & \multicolumn{2}{c}{\textbf{DP-LENS (OFF)}} & \multicolumn{2}{c}{\textbf{DP-LENS (ON)}} \\ \cmidrule(lr){2-3} \cmidrule(l){4-5} 
(Points) & FPS & FT (ms) & FPS & FT (ms) \\ \midrule
200k & 118.97$\pm$3.94 & 8.42 & 114.14$\pm$10.29 & 8.85 \\
500k & 118.75$\pm$4.47 & 8.44 & 113.30$\pm$11.08 & 8.93 \\
800k & 118.91$\pm$2.36 & 8.41 & 112.70$\pm$11.74 & 8.99 \\
1M   & 118.70$\pm$5.12 & 8.45 & 111.78$\pm$12.93 & 9.09 \\
1.5M & 118.90$\pm$4.54 & 8.43 & 110.68$\pm$15.03 & 9.25 \\ \bottomrule
\end{tabular}%
}
\end{table}

\section{User Study 1: Interaction Techniques vs. Occlusion Complexity}

To comprehensively evaluate the effectiveness of the proposed DP-LENS in exploring dense 3D datasets, we conducted a $3 \times 3$ within-subjects design. The two independent variables were the interaction technique (WIM-NAV, V-SLICE, and DP-LENS) and the Topological Occlusion Complexity (detailed below). 

\subsection{Independent Variables}

\paragraph{Interaction Techniques}

\begin{figure}[tb]
 \centering 
 \includegraphics[width=\columnwidth, alt={Visual comparison of three 3D interaction techniques for occlusion management on a vascular dataset: DP-LENS, V-SLICE, and WIM-NAV.}]{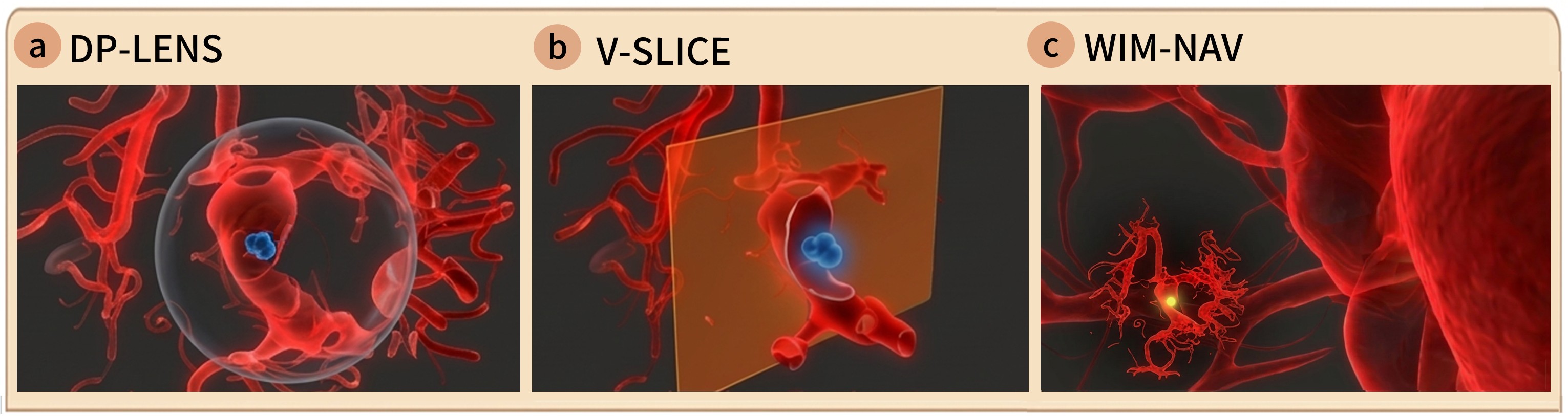}
\caption{3D interaction techniques for occlusion management on the Vessel dataset: (a) DP-LENS, (b) V-SLICE, and (c) WIM-NAV.}
\vspace{-2mm}
 \label{fig:techniques_comparison}
\end{figure}

The primary independent variable was the interaction technique used for visual exploration. We compared our proposed method with two baseline techniques that represent common paradigms for 3D navigation and occlusion management. A visual comparison of the three techniques on the Vessel dataset is shown in \cref{fig:techniques_comparison}. 

\textbf{WIM-NAV (Standard VR Navigation):} This baseline implements the classic WIM interface, where users navigate by scaling and repositioning a synchronized miniature 3D map attached to their dominant hand. The user position is indicated in the miniature view (\cref{fig:techniques_comparison}c). WIM-NAV combines overview+detail and zooming~\cite{Cockburn2009}, and corresponds to the Multiple Viewports strategy in Elmqvist et al.'s occlusion management taxonomy~\cite{Elmqvist2008}. We selected this technique as a baseline for spatial navigation in VR.

\textbf{V-SLICE (Volumetric Slicing):} This baseline uses an interactive planar clipping tool attached to the dominant hand. Users sweep the clipping plane through the dataset to remove occluding foreground structures and reveal internal targets (\cref{fig:techniques_comparison}b). 
This technique corresponds to filtering~\cite{Cockburn2009} and to the Virtual X-Ray strategy~\cite{Elmqvist2008}, and is widely used in scientific and medical visualization~\cite{Weiskopf2003,Preim2013}. We included it as a baseline for occlusion removal.

\textbf{DP-LENS (Proposed):} Our method provides a density-aware volumetric probe for context-preserving exploration. DP-LENS uses polyfocal fisheye deformation and semi-transparency to reveal internal targets while preserving surrounding topology (\cref{fig:techniques_comparison}a). The technique is grounded in the Focus+Context paradigm~\cite{Cockburn2009} and combines properties of Volumetric Probes and Virtual X-Ray~\cite{Elmqvist2008}. In Study 1, DP-LENS operated under manual tracking only, so that the comparison focused on geometric occlusion management performance rather than navigation assistance.

\paragraph{Topological Occlusion Complexity}

\begin{figure}[tb]
 \centering 
 \includegraphics[width=.9\columnwidth, alt={Visualizations of three 3D datasets mapping to different occlusion levels: PointCloud (Level 2), Vortex (Level 3), and Vessel (Level 5).}]{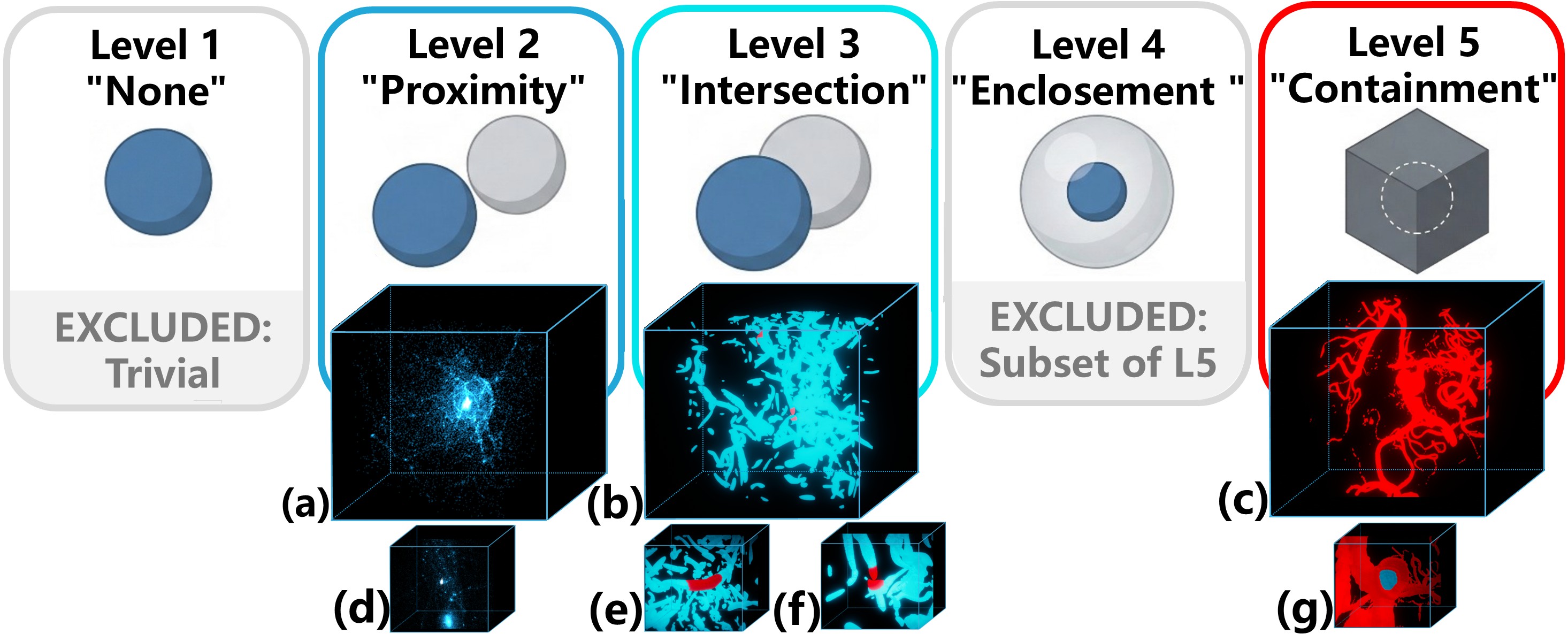}
\caption{3D occlusion taxonomy~\cite{Elmqvist2008} mapped to datasets: (a, d) PointCloud, Level 2; (b, e, f) Vortex, Level 3; (c, g) Vessel, Level 5. Levels 1, 4 excluded (trivial/redundant).}
\vspace{-2mm}
 \label{fig:datasets}
\end{figure}

The second independent variable was topological occlusion complexity. Following the taxonomy of Elmqvist et al.~\cite{Elmqvist2008}, we selected datasets corresponding to three occlusion levels: Proximity (Level 2), Intersection (Level 3), and Containment (Level 5), as summarized in \cref{fig:datasets}. Level 1 was excluded because it involves no meaningful occlusion challenge. Level 4 was not included because Level 5 represents a stricter form of enclosure and subsumes the visibility demands of hollow containment. 


\textbf{PointCloud Task (Level 2: Proximity)}: Participants explored a synthetic dense dataset of exactly 200,000 points, adapted from the benchmark repository of Zhao et al.~\cite{Zhao2024}. The task was to locate and select two high-density white ellipsoidal clusters hidden within a translucent blue point cloud (\cref{fig:datasets}a \& d). The trial ended after both targets were successfully hit.

\textbf{Vortex Task (Level 3: Intersection)}: Participants inspected a fluid-flow dataset resampled to 200,000 particles from the ``Isotropic Turbulence'' enstrophy volume in Open SciVis Datasets~\cite{OpenSciVis}, extracted at an isovalue of 4.5. The task was to trace intertwined vortex structures and distinguish a broken junction from a continuous segment, highlighted in black (\cref{fig:datasets}b, e \& f).

\textbf{Vessel Task (Level 5: Containment)}: Participants explored a vascular dataset resampled to 200,000 volumetric points from the ``Aneurism'' C-arm X-ray scan in Open SciVis Datasets~\cite{OpenSciVis}, extracted at an isovalue of 50. The task was to locate and select two blue blockage targets fully enclosed by vessel walls (\cref{fig:datasets}c \& g). The trial ended after both targets were successfully hit.

\subsection{Task Environment and Design}
\label{sec:us1_task}

To isolate 3D occlusion from the confounding effects of physical locomotion, all datasets were normalized to a 0.6m $\times$ 0.6m $\times$ 0.6m bounding box and positioned at a table height in VR. This setup allowed participants to remain largely stationary and kept the evaluation focused on 3D manual manipulation, pathfinding, and occlusion management within arm's reach.


Participants interacted using Meta Quest 3 controllers. Object selection was performed via ray casting with a virtual laser emitted from the non-dominant controller and confirmed by trigger pull. We chose this input scheme because controller-based ray casting is a standard and well-validated technique for 3D interaction in immersive environments~\cite{Xu2025, Li2025}.

To reduce spatial-memory confounds, we generated three unique target configurations for each dataset. Because each participant completed all three interaction techniques on all three datasets (9 trials total), the target configurations were counterbalanced across techniques such that no participant encountered the same target location twice within a dataset. Each trial had a 3-minute time limit, and participants were instructed to complete the task as quickly and accurately as possible.

\subsection{Hypotheses}
We formulated the following hypotheses based on the trade-offs among 3D occlusion management paradigms identified by Elmqvist et al.~\cite{Elmqvist2008} and Cockburn et al.~\cite{Cockburn2009}: 

\textbf{[H1] Exploration Efficiency}: Compared with WIM-NAV and V-SLICE, DP-LENS integrates focus and context within a single workspace while preserving local topology. We therefore hypothesize that participants will complete occlusion-intensive tasks faster and report higher perceived performance with DP-LENS.

\textbf{[H2] Cognitive Load}: WIM-NAV requires mental integration across separate views, whereas V-SLICE requires users to reconstruct structures disrupted by destructive filtering. Because DP-LENS preserves contextual continuity during local inspection~\cite{Elmqvist2008,Cockburn2009}, we hypothesize that it will reduce mental demand and overall effort relative to both baselines.

\textbf{[H3] Ergonomic Comfort and Fatigue}: Although active volumetric probes often require high-DOF manipulation and can increase physical burden~\cite{besanccon2021state,yu2010fi3d}, DP-LENS augments manual control with magnetic snapping and adaptive deformation. We therefore hypothesize that its physical demand and fatigue will remain low and comparable to V-SLICE, despite the greater topological complexity of the tasks.

\subsection{Participants, Procedure, \& Measurement}

\paragraph{Participants} 
Eighteen participants (11 males, 7 females; mean age = $26.4$, $\mathrm{SD} = 6.45$) were recruited from a university community. All were right-handed, had normal or corrected-to-normal vision and normal color vision, and reported no severe susceptibility to cybersickness. Self-reported VR proficiency, measured on a 7-point Likert scale ($1 = \text{none}$, $7 = \text{expert}$), was moderate on average ($M = 3.17$, $SD = 2.07$). Both User Studies 1 and 2 were approved by the university's Institutional Review Board.

\paragraph{Design and Procedure} 
The study used a within-subjects design. The order of the three interaction techniques (DP-LENS, V-SLICE, and WIM-NAV) and the three datasets (PointCloud, Vortex, and Vessel) was counterbalanced across participants using a Latin square.
Each session lasted approximately 60 minutes. After providing consent and completing a demographic questionnaire, participants completed a 10-minute training session using a separate practice dataset. During training, the experimenter introduced all three techniques, and participants practiced navigation, ray-casting selection, and trigger-based target confirmation until they were familiar with the controls.
In the formal evaluation phase, each participant completed 9 trials (3 techniques $\times$ 3 datasets). The 9 trials were organized into three technique blocks to avoid frequent switching between control schemes. Within each block, participants completed the three dataset conditions. After each technique block, participants removed the headset and completed the NASA Task Load Index (NASA-TLX)~\cite{hart1988development}, Borg Rating of Perceived Exertion (Borg RPE)~\cite{borg1998borg}, and Fast Motion Sickness Scale (FMS) Sickness~\cite{keshavarz2011validating} questionnaires. A mandatory 1-minute break was provided between blocks.

\paragraph{Measurements} 
Objective performance was quantified using Task Completion Time, recorded from trial onset to successful target selection. Trials that exceeded the time limit were marked as failures. Subjective measures included NASA-TLX for workload, Borg RPE for perceived physical exertion, and FMS for simulator sickness. We also collected user preference rankings and think-aloud comments to contextualize participants’ strategies and experiences.

\paragraph{Statistical Analysis}
Quantitative data were analyzed using repeated-measures statistical tests. Normality was assessed with the Shapiro--Wilk test. For normally distributed outcomes \textcolor{black}{($p > .05$)}, we used one-way repeated-measures ANOVA; sphericity was evaluated with Mauchly’s test, and Greenhouse--Geisser correction was applied when necessary. For non-normal data \textcolor{black}{($p < .05$)}, we used Friedman tests. Significant main effects were followed by Bonferroni-corrected pairwise comparisons using paired t-tests for parametric data and Wilcoxon signed-rank tests for nonparametric data. 

\subsection{Results}

\begin{figure}[h]
    \centering
    \includegraphics[width=0.6\linewidth, alt={Bar charts showing task completion time and success rates for User Study 1 across PointCloud, Vortex, and Vessel datasets.}]{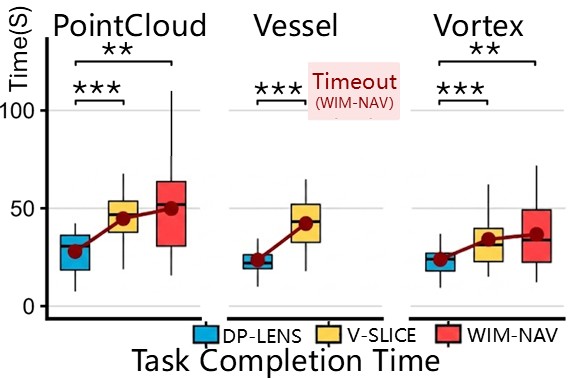}
    \caption{Results of User Study 1 -- Task Performance and Efficiency. (**$p < .01$, ***$p < .001$).}
    \label{fig:time-us1}
\end{figure}

\paragraph{Task Performance and Efficiency} 
Task completion rates and completion times are visually summarized in \cref{fig:time-us1}, with detailed descriptive statistics provided in Table \ref{tab:study1_results}. For the Vortex and PointCloud datasets, all participants completed the task successfully under all three techniques. In the Vessel dataset, however, WIM-NAV yielded a 0\% completion rate, with all participants reaching the trial timeout, whereas both V-SLICE and DP-LENS maintained a 100\% completion rate. WIM-NAV was therefore excluded from the completion-time analysis for the Vessel dataset.

For completion time, DP-LENS consistently outperformed the comparison techniques (see Table \ref{tab:study1_results}). In the Vortex dataset, RM-ANOVA showed a significant main effect of the interaction technique. As Mauchly's test indicated a violation of sphericity ($p < .001$), the Greenhouse-Geisser correction was applied ($F(2, 34) = 5.26, p = .026$). Bonferroni-correction pairwise comparisons showed that DP-LENS was significantly faster than both V-SLICE ($p = .0002$) and WIM-NAV ($p = .0096$). In the PointCloud dataset, the sphericity assumption was satisfied ($p > .05$), and RM-ANOVA again revealed a significant main effect of technique ($F(2, 34) = 8.89, p = .0008$). Participants completed tasks significantly faster using DP-LENS compared to both V-SLICE ($p = .0008$) and WIM-NAV ($p = .0037$). Finally, for the Vessel dataset, a paired t-test showed that DP-LENS was significantly faster than V-SLICE ($p < .0001$).


\begin{figure}[t]
    \centering
    \includegraphics[width=\linewidth, alt={Bar charts displaying NASA-TLX workload scores including Mental Demand, Physical Demand, and Effort for the three interaction techniques.}]{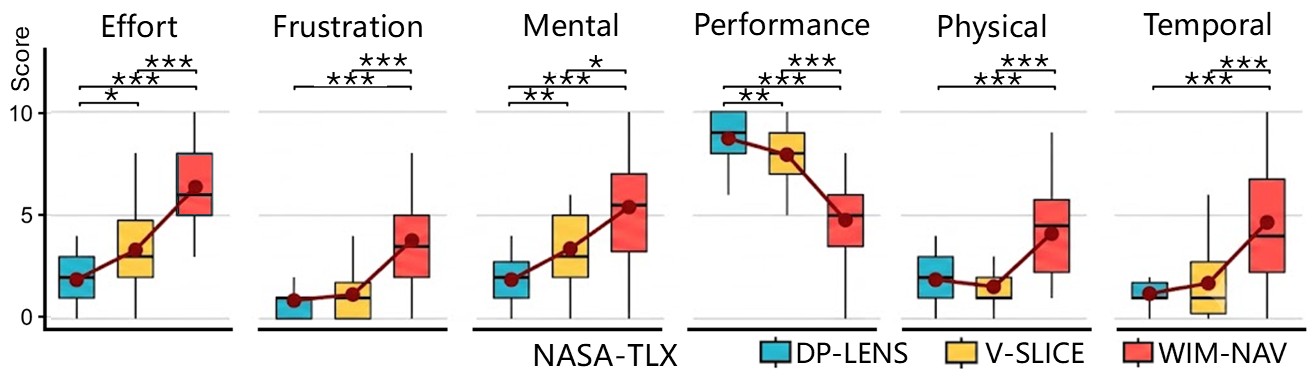}
    \caption{Results of User Study 1 -- user workload measured by the NASA-TLX questionnaire (*$p < .05$, **$p < .01$, ***$p < .001$).}
    \label{fig:nasa-tlx-1}
\end{figure}

\paragraph{Cognitive Load, Performance, and Effort} 
Subjective workload results are shown in \cref{fig:nasa-tlx-1} and detailed in Table \ref{tab:study1_results}. When evaluating the overall NASA-TLX workload (calculated as the average of the six subscales, with the Performance score inverted), statistical analysis revealed a significant main effect of the interaction technique ($p < .001$). DP-LENS yielded a significantly lower overall workload compared to both V-SLICE and WIM-NAV (both $p < .01$). For the individual dimensions, the Friedman test revealed a significant main effect of technique for Mental Demand ($p = .0001$). Bonferroni-corrected post-hoc comparisons confirmed that DP-LENS imposed significantly lower mental demand than both V-SLICE ($p = .0059$) and WIM-NAV ($p = .0009$). A corresponding pattern was observed for perceived performance: participants reported significantly higher NASA-TLX Performance scores with DP-LENS than with V-SLICE ($p = .0071$) and WIM-NAV ($p = .0003$). For overall Effort, the data satisfied normality and sphericity assumptions. RM-ANOVA showed a significant main effect of technique ($F(2, 34) = 49.47, p < .0001$). Bonferroni-corrected Post-hoc tests showed that DP-LENS required significantly less effort than V-SLICE ($p = .0137$) and WIM-NAV ($p < .0001$).

\begin{figure}[b]
    \centering
    \includegraphics[width=0.8\linewidth, alt={Bar charts showing Borg RPE scores for arm and neck fatigue, and FMS Sickness scores for the three interaction techniques.}]{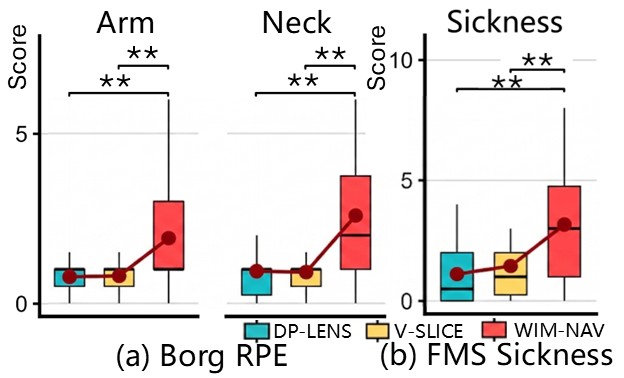}
    \caption{Results of User Study 1 -- Borg RPE and FMS Sickness (**$p < .01$).}
    \label{fig:sickness-1}
\end{figure}

\paragraph{Fatigue and Motion Sickness} 
As shown in \cref{fig:sickness-1} and Table \ref{tab:study1_results}, Borg RPE ratings revealed significant differences in both arm fatigue ($p = .0002$) and neck fatigue ($p = .0005$). In both cases, DP-LENS and V-SLICE produced significantly lower fatigue than WIM-NAV. Similarly, FMS scores showed a significant main effect for motion sickness ($p = .0001$), where both DP-LENS and V-SLICE resulted in significantly lower sickness scores than WIM-NAV.

\begin{table}[tb]
\centering
\caption{Summary of statistical results for User Study 1. Values are presented as Mean $\pm$ Standard Deviation (SD).}
\label{tab:study1_results}
\resizebox{\columnwidth}{!}{%
\begin{tabular}{@{}lccc@{}}
\toprule
\textbf{Metric} & \textbf{DP-LENS} & \textbf{V-SLICE} & \textbf{WIM-NAV} \\ \midrule
\multicolumn{4}{l}{\textit{Task Completion Time (Unit: Seconds)}} \\
\quad PointCloud (Level 2) & \textbf{28.31 $\pm$ 10.23} & 44.81 $\pm$ 16.08 & 49.86 $\pm$ 24.06 \\
\quad Vortex (Level 3) & \textbf{24.35 $\pm$ 10.31} & 34.38 $\pm$ 15.61 & 36.88 $\pm$ 16.78 \\
\quad Vessel (Level 5) & \textbf{24.01 $\pm$ 8.88} & 42.33 $\pm$ 13.07 & Timeout (0\%) \\ \midrule
\multicolumn{4}{l}{\textit{Subjective Workload (Scale: 0-10, Lower is better except Performance)}} \\
\quad Overall Workload & \textbf{1.51 $\pm$ 0.72} & 2.20 $\pm$ 1.13 & 4.91 $\pm$ 1.59 \\
\quad Mental Demand & \textbf{1.89 $\pm$ 1.13} & 3.39 $\pm$ 1.91 & 5.39 $\pm$ 3.03 \\
\quad Effort & \textbf{1.89 $\pm$ 1.13} & 3.33 $\pm$ 2.14 & 6.28 $\pm$ 2.02 \\
\quad Performance & \textbf{8.72 $\pm$ 1.18} & 7.94 $\pm$ 1.43 & 4.78 $\pm$ 2.05 \\ \midrule
\multicolumn{4}{l}{\textit{Physical Fatigue \& Sickness (Scale: 0-10, Lower is better)}} \\
\quad Borg RPE: Arm Fatigue & \textbf{0.78 $\pm$ 0.49} & 0.81 $\pm$ 0.46 & 1.92 $\pm$ 1.57 \\
\quad Borg RPE: Neck Fatigue & 0.94 $\pm$ 0.73 & \textbf{0.92 $\pm$ 0.73} & 2.58 $\pm$ 2.17 \\
\quad FMS Simulator Sickness & \textbf{1.11 $\pm$ 1.37} & 1.44 $\pm$ 1.34 & 3.17 $\pm$ 2.50 \\ \bottomrule
\end{tabular}%
}
\end{table}

\paragraph{User Preferences and Qualitative Feedback} 
Preference rankings are summarized across the three datasets. DP-LENS was the most preferred technique in all cases. In the Vortex dataset, 11 participants (61.1\%) preferred DP-LENS, compared with 5 (27.8\%) for V-SLICE and 2 (11.1\%) for WIM-NAV; this distribution differed significantly from chance ($\chi^2(2)=7.00$, $p=.030$). In the PointCloud dataset, 14 participants (77.8\%) preferred DP-LENS, 4 (22.2\%) preferred V-SLICE, and none preferred WIM-NAV ($\chi^2(2)=17.33$, $p<.001$).
In the Vessel dataset, 17 out of 18 participants(94.4\%) preferred DP-LENS, with the remaining participant selecting V-SLICE and none selecting WIM-NAV ($\chi^2(2) = 30.33, p < .001$).

Participants considered V-SLICE acceptable for lower-occlusion PointCloud and Vortex tasks, but reported increasing difficulty in the Level 5 Vessel condition. One participant noted: \textit{``Using volume slicing in the heart vessels was troublesome because I had to keep pushing the slice back and forth; I even missed a target on my first pass.''} This aligns with the higher Mental Demand scores observed for V-SLICE. Participants also reported that WIM-NAV became difficult to use in high-occlusion structures. As one user stated: \textit{``[In the PointCloud and Vortex tasks], it was not that hard; you can find the targets by just looking. But in the heart [Vessel], it was too difficult... it's nearly impossible to find anything using that minimap way.''} In contrast, participants described DP-LENS as providing local details while maintaining global awareness. One participant summarized this advantage as follows: \textit{``This [technique] is so clear... [I] can find [the targets] instantly, it's not tiring at all.''} 

\paragraph{Preliminary Expert Feedback}
A chief physician from a tertiary hospital reviewed the three interaction paradigms using the Vessel dataset. The expert noted that WIM-NAV's miniature map suffered from visual clutter: \textit{``That [WIM-NAV]... the minimap... it is just too cluttered. Those star-like [vascular noise] clusters... you cannot see any detail at all. It is just not usable in a clinical setting''}. Furthermore, while V-SLICE felt familiar, it disrupted spatial continuity: \textit{``It [V-SLICE] feels like how we normally look at [CT] scans, but the biggest problem is that it lacks spatial continuity between adjacent slices''}. In contrast, the expert preferred DP-LENS, observing that its ``perspective effect" revealed multiple layers simultaneously. The expert suggested this could hold potential for workflows such as preoperative planning, stating: \textit{``This [DP-LENS] is fast... being able to see multiple layers and points of interest at once... I prefer this [DP-LENS] approach.''}.

\subsection{Discussion}

The first study supports our hypotheses on 3D occlusion management and clarifies how different interaction paradigms respond to increasing topological complexity. 

\paragraph{Exploration Efficiency} 
\textit{H1} predicted that DP-LENS would enable faster and more effective target exploration than WIM-NAV and V-SLICE. The results support this hypothesis. DP-LENS achieved the fastest completion times in all three datasets and maintained a 100\% completion rate even in the most challenging Level 5 Containment condition. By contrast, WIM-NAV failed completely in the Vessel dataset, where all participants reached the trial timeout. Prior research by Cockburn et al.~\cite{Cockburn2009} and Elmqvist et al.~\cite{Elmqvist2008} indicates that while Overview+Detail techniques (e.g., Multiple Viewports / WIM) are foundational for multiscale navigation, they inherently introduce divided attention and visual clutter when miniaturizing dense datasets. Our findings from Study 1 also support this view: Overview+Detail techniques may not be suitable for dense internal structures that require precise path following and fine-grained spatial localization. V-SLICE remained effective in terms of task completion, but was consistently slower than DP-LENS. Together, these findings indicate that integrating focus and context within a single local workspace offers a practical advantage for occlusion-intensive exploration.

\paragraph{Cognitive Load} 

\textit{H2} predicted that DP-LENS would reduce cognitive load relative to the two baseline techniques. The subjective results support this hypothesis. DP-LENS produced significantly lower Mental Demand and Effort scores than both V-SLICE and WIM-NAV. These results are consistent with prior arguments that Overview+Detail requires mental integration across separated views~\cite{Cockburn2009,Elmqvist2008}, whereas destructive slicing removes contextual structure and forces users to reconstruct spatial relationships from partial views~\cite{Elmqvist2008}. Participant feedback further reflected these difficulties, WIM-NAV disrupted spatial orientation in dense internal navigation, while V-SLICE required repeated slice adjustment and increased mental effort. In contrast, the  density-aware Focus+Context design of DP-LENS appears to preserve local detail while maintaining sufficient surrounding context for continuous interpretation, and can reduce the interpretive burden of deep 3D inspection.

\paragraph{Ergonomic Comfort and Fatigue} 
\textit{H3} predicted that DP-LENS would maintain low physical demand and fatigue despite the manipulation requirements of active volumetric probes. The results support this hypothesis. A common critique of active 3D volumetric probes is that their continuous 6-DOF manipulation typically induces significant arm fatigue and requires high manual dexterity~\cite{besanccon2021state,yu2010fi3d}. However, our objective and subjective findings show that Although DP-LENS is based on an active probe metaphor, it did not produce higher arm or neck fatigue than V-SLICE, and both techniques were significantly less fatiguing than WIM-NAV. 
This suggests that the magnetic snapping and adaptive scaling mechanisms reduced the need for continuous high-precision manual adjustment, allowing DP-LENS to improve cognitive outcomes without introducing the ergonomic penalties commonly associated with free 6-DOF probe manipulation.

\paragraph{Summary and Transition to Study 2} 

To sum up, the findings from User Study 1 show that DP-LENS outperformed the comparison techniques for dense desktop-scale occlusion management.
These findings establish its effectiveness for local, within-arm's-reach exploration. However, as immersive analytics increasingly scales to room-scale environments, the ergonomic bottlenecks shift from localized manual dexterity to gross physical locomotion and long-distance arm extension (e.g., the Gorilla Arm effect). This challenge motivates the Hybrid Voice-Initiated Auto-Routing framework evaluate in User Study 2.

\section{User Study 2: Hybrid Voice-Initiated Auto-Routing}

The first study showed that manual DP-LENS supported effective exploration of dense occlusions within a desktop-scale interaction volume.  However, many immersive analytics often requires users to explore expanded datasets at larger physical scales, such as extended vascular structures or large fluid-flow fields. In such settings, manual 6-DOF manipulation may become more demanding because users must traverse a larger workspace and sustain mid-air interaction for longer periods.

To address this scalability challenge, we introduced the Hybrid Voice-Initiated Auto-Routing framework (\cref{sec:ai_nav}), which supplements manual control with LLM-assisted long-range routing. 
User Study 2 evaluates whether this augmentation improves performance and reduces user burden relative to manual DP-LENS across two spatial scales. We tested the following hypotheses: \textbf{[H4]:} The Hybrid Voice-Initiated Auto-Routing will significantly reduce task completion time and accelerate target acquisition compared to manual navigation. \textbf{[H5]:} Hybrid Voice-Initiated Auto-Routing will reduce perceived physical demand and mental workload relative to manual navigation when exploring dense, room-scale environments. \textbf{[H6]:} The effect of dataset scale on task completion time will be smaller under Hybrid Voice-Initiated Auto-Routing than under manual navigation.

\subsection{Participants, Study Design, and Measurement}

\paragraph{Participants and Apparatus} To reduce learning effects and spatial memory carryover from User Study 1, we recruited a new cohort of 16 participants. (10 males, 6 females; mean age = 23.9, $\mathrm{SD} = 4.0$). All participants had normal or corrected-to-normal vision and normal color vision. The hardware setup and procedures matched those used in Study 1.


\paragraph{Task and Dataset} To maintain consistency in occlusion difficulty, we reused the Vessel dataset from User Study 1, which represented the highest occlusion level in our study set. The task was to navigate the 3D space, reveal internal vessel structures using the lens, and acquire two designated blockage targets. Selection was performed using the same controller-based laser raycasting technique as in Study 1, with confirmation by trigger press. Each trial ended after the second target was successfully acquired. We prepared four target sets, each containing two blockage targets with matched occlusion difficulty. These target sets were counterbalanced across conditions to reduce spatial learning effects and target memorization effects. 


\paragraph{Study Design} The study used a $2 \times 2$ within-subjects design with \emph{Navigation Mode} (Manual DP-LENS vs. Hybrid Voice-Initiated Auto-Routing) and \emph{Dataset Scale} (Desktop-scale vs. Room-scale) as independent variables. To evaluate spatial scalability, we focused on the Vessel dataset because it imposed the strongest occlusion challenge in User Study 1. 
The dataset was rendered at two scales: a desktop-scale condition (bounded within a 0.6m $\times$ 0.6m $\times$ 0.6m interaction volume) and a room-scale condition (bounded within a 2.0m $\times$ 2.0m $\times$ 2.0m walkable tracking space). The two-meter condition was chosen to approximate a typical consumer VR room-scale boundary. All participants began each trial seated. In the desktop-scale condition, tasks could generally be completed while seated, whereas the room-scale condition often required participants to stand and reposition themselves within the tracked area. Each participant completed four trials. Condition order and target-set assignment were counterbalanced using a balanced Latin square design~\cite{Lazar2017research}.


\paragraph{Measurement} Task completion time was recorded automatically from trial onset to acquisition of the second target. After completing all trials under a given navigation mode, participants filled out the NASA-TLX for perceived workload, Borg RPE ratings for localized physical fatigue (Arm and Neck), and the FMS scale for simulator sickness. We also collected concurrent think-aloud comments and post-study interviews feedback to capture users' interaction strategies and subjective feedback regarding the two navigation modes.

\begin{figure}[tb]
 \centering 
\includegraphics[width=\columnwidth, alt={Charts illustrating the results of User Study 2, comparing Task Completion Time, Physical Demand, and Mental Demand between Manual and Hybrid Voice-Initiated Auto-Routing modes across two spatial scales.}]{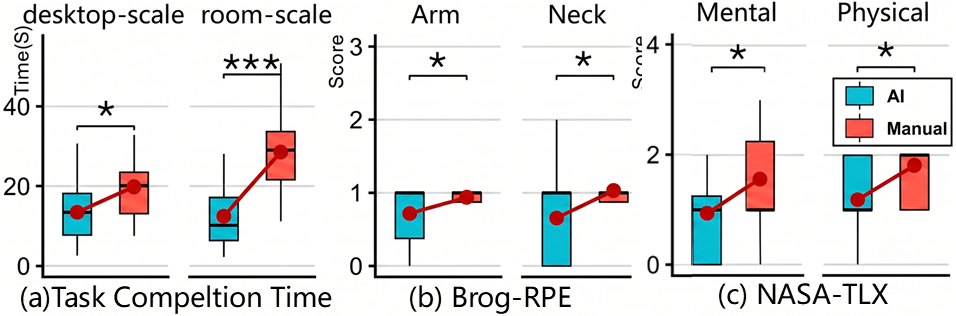}
\caption{Results of User Study 2 (*$p < .05$, ***$p < .001$).}
\vspace{-2mm}
 \label{fig:us2_results}
\end{figure}

\subsection{Results}

\paragraph{Task Completion and Efficiency} 
The task completion rate was 100\% across all conditions in User Study 2, with all participants successfully acquiring the targets. Regarding the exploration time, because Shapiro--Wilk tests indicated non-normality \textcolor{black}{($p < .05$)}, task completion times were log-transformed prior to analysis. A two-way RM-ANOVA showed a significant main effect of the navigation mode on task completion time ($F(1, 15) = 22.17, p < .001$), indicating that participants completed the tasks significantly faster with the auto-routing navigation system compared to the manual navigation (\cref{fig:us2_results}a). We also found a significant interaction between the navigation mode and the dataset scale ($F(1, 15) = 6.68, p < .05$). Post-hoc comparisons showed that completion time increased significantly from the Desktop-scale to the Room-scale condition under manual navigation ($p < .05$), whereas no significant scale-related difference was observed under auto-routing navigation ($p = .54$). Furthermore, to assess the temporal overhead of the voice interface, we analyzed the system logs. The average latency from the end of a voice command to the initiation of the auto-routing lens movement was approximately 2.88s (ASR transcription: $\sim$1.16s, LLM inference: $\sim$1.72s). Because this processing executes asynchronously, it did not cause any frame-rate drops or block the continuous rendering loop.

\paragraph{Subjective Perceived Workload (NASA-TLX)} 
Due to the non-normal distribution of the subjective rating data (Shapiro-Wilk, $p < .05$), a series of Wilcoxon signed-rank tests were performed. As summarized in Table \ref{tab:study2_results}, when evaluating the overall NASA-TLX workload (calculated as the average of the six subscales, with the Performance score inverted), the auto-routing system significantly reduced the total cognitive burden compared to manual control ($p < .05$). Specifically for the individual subscales, the results showed a significant reduction in mental demand ($V=16.5, Z=-2.11, p=.032, r=0.37$, \cref{fig:us2_results}c) under the auto-routing navigation compared to the manual control. No significant differences were found between the two conditions for temporal demand ($V=15.5, Z=-0.14, p=.719, r=0.02$), performance ($V=28.5, Z=0.38, p=.916, r=0.07$), effort ($V=40.5, Z=0.08, p=.905, r=0.01$), and frustration ($V=17.0, Z=-1.21, p=.265, r=0.21$).

\paragraph{Physical Demand \& Ergonomics} We also found significant effects of the navigation mode for overall physical demand ratings ($V=10.0, Z=-2.18, p=.029, r=0.39$), Borg arm fatigue ($V=0.0, Z=-2.07, p=.038, r=0.37$), and Borg neck fatigue ($V=0.0, Z=-2.23, p=.026, r=0.39$), all detailed in Table \ref{tab:study2_results}. Observationally, tracing targets in room-scale datasets under manual control frequently required sustained arm elevation and repetitive head rotations. In contrast, the Hybrid Voice-Initiated Auto-Routing system yielded significantly lower physical exertion scores.

\paragraph{Simulator Sickness Questionnaire (FMS)} 
Similar to User Study 1, we analyzed the motion sickness scores. The mean FMS scores for both the auto-routing condition and the manual condition remained comparably low (see Table \ref{tab:study2_results}), and there was no significant difference between the two techniques ($p = .16$).

\begin{table}[tb]
\centering
\caption{Summary of subjective results for User Study 2 (Scale: 0-10, Lower is better). Values are presented as Mean $\pm$ Standard Deviation (SD).}
\label{tab:study2_results}
\begin{tabular}{@{}lcc@{}}
\toprule
\textbf{Metric} & \textbf{Hybrid Voice} & \textbf{Manual Control} \\ \midrule
\multicolumn{3}{l}{\textit{Subjective Workload (NASA-TLX)}} \\
\quad Overall Workload & \textbf{1.32 $\pm$ 0.71} & 1.57 $\pm$ 0.70 \\
\quad Mental Demand & \textbf{0.94 $\pm$ 0.77} & 1.56 $\pm$ 0.96 \\ \midrule
\multicolumn{3}{l}{\textit{Physical Fatigue \& Sickness}} \\
\quad Overall Physical Demand & \textbf{1.19 $\pm$ 0.66} & 1.81 $\pm$ 0.98 \\
\quad Borg RPE: Arm Fatigue & \textbf{0.72 $\pm$ 0.45} & 0.94 $\pm$ 0.54 \\
\quad Borg RPE: Neck Fatigue & \textbf{0.66 $\pm$ 0.60} & 1.03 $\pm$ 0.78 \\
\quad FMS Simulator Sickness & \textbf{0.97 $\pm$ 0.97} & 1.25 $\pm$ 1.13 \\ \bottomrule
\end{tabular}
\end{table}

\paragraph{User Feedback and Behavioral Divergence} The qualitative data suggest that participants used the Hybrid Voice-Initiated Auto-Routing system differently depending on the navigation scale and task granularity. When navigating the room-scale condition, several participants reported that voice-based navigation helped them avoid unnecessary physical locomotion. For example, one participant noted: \textit{``I can just tell it [the AI] to move the magnifier over there, and I don't have to move [my body] at all.''} 
At the same time, some participants noted occasional mismatches between their verbal intent and the system response. One participant commented:\textit{``I told it to go to the left vessel, but it directly showed me [the area on the] bottom-left, so I still had to manually move it [the lens] a bit.''}

In contrast, during fine-grained adjustment in the desktop-scale condition, some participants preferred manual input because of the latency associated with speech interaction. As one participant stated: \textit{``Language always needs time to process and react... manual control is much faster.''} Consequently, they preferred immediate manual correction over issuing a subsequent voice command, explaining: \textit{``I prefer to use my hands for the adjustment rather than talking again, unless my hands are occupied.''} Despite this tradeoff, 9 of the 16 participants chose to rely exclusively on voice commands throughout the task. When asked about this strategy during the interview, particularly for targets already within arm's reach, one user confessed: \textit{``When it's close, feeling-wise, using hands might be faster, but voice is [just more] convenient; I don't have to move [my hands at all].''} This sentiment was widely shared, indicating that for many users, the perceived convenience and reduced physical effort of the voice command system easily outweighed the absolute speed advantage of direct manual manipulation.

\subsection{Discussion}

\paragraph{Mitigating Ergonomic Bottlenecks in Room-Scale VR} Prior HCI research consistently identifies that prolonged, unsupported 6-DOF manual manipulation in mid-air environments inevitably induces severe physical fatigue, widely recognized as the ``Gorilla Arm'' effect~\cite{besanccon2021state,yu2010fi3d}. Our Borg RPE and physical demand data firmly support this premise for traditional manual control, as physical fatigue escalated significantly when datasets expanded to room-scale. However, our findings show that the Hybrid Voice-Initiated Auto-Routing system successfully suppressed this fatigue surge ($p < .05$), fully supporting \textit{H5}. By delegating macro-spatial routing to the LLM module, the system significantly reduces the user's physical payload. Rather than executing continuous, high-amplitude shoulder and arm movements to track complex topologies, users only need to issue a verbal intent, wait for the automated lens flight, and perform minor manual micro-adjustments if necessary. This workflow effectively bypasses the primary ergonomic bottlenecks of large-scale immersive analytics.

\paragraph{Decoupling Interaction Cost from Spatial Constraints} Previous studies on spatial navigation emphasize that exploring dense, featureless 3D environments inherently inflates navigation friction and cognitive disorientation (the ``desert fog'' effect)~\cite{Julier2000,Cockburn2009}. As physical scale increases, the temporal and cognitive costs of spatial pathfinding escalate significantly. Our significant interaction effect ($p < .05$) directly illustrates this boundary: manual navigation time severely degraded as the dataset expanded. Conversely, the temporal cost of the LLM-assisted condition remained remarkably flat ($p = .54$) across the tested spatial scales, supporting \textit{H4} and \textit{H6}. This spatial decoupling occurs because the LLM's semantic parsing translates absolute physical 3D distance into a localized region selection. Consequently, the interaction efficiency is less bound by the physical dimensions of the virtual world, making it highly scalable within standard room-scale environments.

\paragraph{The Principle of Least Effort and Automation Reliance} Research in human-automation interaction and behavioral psychology suggests that humans generally adhere to the \textit{Principle of Least Effort} ~\cite{zipf2016human}; when automated systems reach an acceptable threshold of reliability, users exhibit ``automation reliance,'' voluntarily trading potential speed for reduced physical and cognitive exertion~\cite{parasuraman1997humans}. Our observational data aligns with this theory. Despite the hybrid system's allowance for fast manual micro-adjustments, a majority of users (9/16) exhibited strong automation reliance, choosing to exclusively use voice commands. As reflected in their feedback (e.g., \textit{``voice is convenient, I don't have to move''}) and the significantly lower NASA-TLX mental demand scores, the ergonomic payoff was highly prioritized. This reveals a compelling user behavior in immersive analytics: when the system's accuracy and latency meet baseline expectations, the cost of waiting for natural language processing is perceived as far lower than the cost of physical arm movement. Ultimately, users are willing to sacrifice absolute interaction speed in exchange for a low-fatigue, highly automated analytical experience.


\section{KEY IMPLICATIONS AND LIMITATIONS}\label{sec:discuss}

At its core, our research set out to address a very user-centric XR  problem: the physical fatigue and cognitive strain often associated with exploring complex 3D environments. While our studies demonstrated that the DP-LENS framework can make this process significantly faster and more comfortable, some of the most fascinating observations emerged when we looked beyond the raw performance metrics. By examining how participants chose to interact alongside the AI, we noted intriguing patterns in user behavior. The way users navigated the trade-off between physical effort and conversational AI latency challenges certain traditional design assumptions, offering nuanced perspectives on automation reliance that can help inform the future development of immersive analytics and 3D interfaces.

\paragraph{Behavioral Divergence: Automation Reliance vs. Direct Manipulation} Foundational HCI and immersive analytics research often emphasizes the benefits of ``direct manipulation,'' where continuous control and immediate visual feedback help users maintain spatial ownership and interaction flow~\cite{besanccon2021state, Cockburn2009}. Based on these well-established principles, we initially anticipated a strict hybrid division of labor: we hypothesized that users would leverage the LLM agent solely for macro-traversal, but would actively reject the LLM's inherent conversational latency during micro-adjustments, reverting to direct manual control when targets were within arm's reach. 

However, our observational data revealed a fascinating behavioral divergence. A majority of participants subverted our expectation by actively abandoning direct manipulation, choosing to rely exclusively on voice commands even for immediate peripersonal targets. As one user noted, the latency of waiting for the AI was a highly worthwhile trade-off for the convenience of physical immobility. For these users, the ergonomic cost of physical movement heavily outweighed the temporal cost of conversational latency; once the voice-initiated routing system's reliability crossed a certain threshold, they exhibited strong automation reliance, willingly sacrificing the immediacy of direct control for a low-effort experience. 

On the other hand, a subset of users exhibited the exact trade-off we initially predicted. When targets were already close by, the inherent latency of conversational AI disrupted their interaction flow, leading them to strongly prefer direct manual manipulation for fine-grained micro-adjustments. They felt that fully replacing manual dexterity with automation reduced their interaction immediacy and spatial ownership. This interesting dichotomy highlights that while voice assistance excels at macro-exploration, completely replacing manual operation is not universally optimal, echoing the necessity for a balanced human-AI collaborative design.

\paragraph{Design Implications} Our studies generate design clues with three implications for immersive visual analytics and LLM-assisted navigation in VR environments. 

\textit{DI1: Implement Intent-Driven Systems for Macro-Exploration.} As the size of the virtual environment or dataset increases, continuous manual navigation induces severe physical fatigue. System designers could prioritize integrating Automated trajectory planning to automate long-distance traversal, thereby minimizing user movement and preserving physical stamina. 

\textit{DI2: Adapt to Task Proximity (Support Hybrid Modes).} For simple micro-adjustments within the user's peripersonal space, LLM latency might introduce redundant latency. Therefore, interaction designs could benefit from offering hybrid control; users could retain the ability to seamlessly transition to direct manual control for localized, fine-grained interactions rather than relying solely on automation. 

\textit{DI3: Prioritize Context-Preserving Deformation.} In complex tasks involving dense volumetric data, destructive techniques like slicing increase cognitive load. Systems are encouraged to favor non-destructive, density-aware deformations (similar to DP-LENS) to allow internal inspection while maintaining peripheral spatial awareness.

\paragraph{Limitations and Future Work} First, the reliance on verbal communication introduces processing latency that can disrupt rapid micro-adjustments. Future systems should explore multimodal intent recognition (e.g., eye-tracking plus gestures) to reduce conversational overhead. Second, our datasets were static; updating density-aware fields for dynamic topologies presents a computational challenge. Third, to better evaluate the time-fatigue trade-off, future studies could implement a `busy' XR dual-task scenario (e.g., searching for information with AI while simultaneously performing a manual data input task) to observe how users dynamically jump between AI assistance and manual control. Finally, while our current framework was evaluated exclusively in VR, future work should investigate how DP-LENS adapts to the unique optical and physical occlusion challenges of Augmented Reality (AR). Additionally, while our study included initial feedback from a single senior medical expert, the primary participant pool consisted of university students.


\section{Conclusion} 
In this paper, we presented DP-LENS, an interaction framework equipped with topology-driven auto-routing, designed to help address severe topological occlusions while mitigating spatial cognitive load and physical fatigue in immersive 3D visual analytics. By utilizing a density-aware volumetric fisheye lens combined with context-preserving X-ray rendering, our findings indicate that maintaining continuous spatial topology can play a key role in reducing users' cognitive workload during deep internal inspections. Furthermore, our research suggests the practical potential of deploying Large Language Models as supplementary, voice-based target selection tools in immersive environments. By feeding the user's semantic intent into the topology-driven auto-routing algorithm, our system effectively decouples human interaction costs from the physical scale of the dataset, facilitating less physically demanding exploration. Ultimately, our work suggests that providing a synergistic control spectrum—algorithm-driven macro-routing combined with the option for direct manual micro-adjustments in peripersonal space—accommodates diverse user preferences, balancing the observed user inclination toward low-effort automation with the need for immediate, fine-grained control.

\acknowledgments{
This research is supported by the Hong Kong Polytechnic University (PolyU)'s Start-up Fund for New Recruits Under Grant (Project ID: P0046056), PolyU Department of Industrial and Systems Engineering Under Grant (Project ID: P0056354), PolyU Faculty of Engineering Under Grant (Project ID: P0064048), University of Warwick – The Hong Kong Polytechnic University Joint Seed Fund Under Grant (Project ID: P0063938), the Research Grant Council's General Research Fund Under Grant (Project ID: P0056902), and PolyU RIAM -- Research Institute for Advanced Manufacturing Under Grant (No. P0056767, P0064368). The authors also acknowledge the use of GPT-5.2 (OpenAI) for language checking and polishing, and Gemini 3.1 Flash (Google) for illustrative images, elements or materials.
}

\bibliographystyle{abbrv-doi-hyperref}

\bibliography{template}

\begin{thebibliography}{10}

\bibitem{Ahn2022}
M.~Ahn et~al.
\newblock Do as i can, not as i say: Grounding language in robotic affordances.
\newblock {\em arXiv preprint arXiv:2204.01691}, 2022.

\bibitem{bauer2021multi}
D.~Bauer, C.~Zheng, O.-H. Kwon, and K.-L. Ma.
\newblock A multi-layout design for immersive visualization of network data.
\newblock {\em arXiv preprint arXiv:2112.10272}, 2021.

\bibitem{Bermejo2021Button}
C.~Bermejo, L.~H. Lee, P.~Chojecki, D.~Przewozny, and P.~Hui.
\newblock Exploring button designs for mid-air interaction in virtual reality: A hexa-metric evaluation of key representations and multi-modal cues.
\newblock {\em Proceedings of the ACM on Human-Computer Interaction}, 5(EICS):1--26, 2021.

\bibitem{besanccon2021state}
L.~Besan{\c{c}}on, A.~Ynnerman, D.~F. Keefe, L.~Yu, and T.~Isenberg.
\newblock The state of the art of spatial interfaces for 3d visualization.
\newblock In {\em Computer Graphics Forum}, vol.~40, pp. 293--326. Wiley Online Library, 2021.

\bibitem{borg1998borg}
G.~Borg.
\newblock {\em Borg's perceived exertion and pain scales.}
\newblock Human kinetics, 1998.

\bibitem{Carpendale1997}
M.~S.~T. Carpendale et~al.
\newblock Distortion viewing techniques for 3-dimensional data.
\newblock In {\em Proceedings of the IEEE Symposium on Information Visualization}, pp. 46--53, 1997.

\bibitem{Cautun2013}
M.~Cautun, R.~van~de Weygaert, and B.~J.~T. Jones.
\newblock Nexus: tracing the cosmic web connection.
\newblock {\em Monthly Notices of the Royal Astronomical Society}, 429(2):1286--1308, 2013.

\bibitem{Chen2008}
L.~Chen and I.~Fujishiro.
\newblock Optimizing parallel performance of streamline visualization for large distributed flow datasets.
\newblock In {\em 2008 IEEE Pacific Visualization Symposium}, pp. 87--94. IEEE, 2008.

\bibitem{Qwen2024Audio}
Y.~Chu, J.~Xu, Q.~Yang, H.~Wei, X.~Wei, Z.~Guo, et~al.
\newblock {Qwen2-Audio Technical Report}.
\newblock {\em arXiv preprint arXiv:2407.10759}, 2024.

\bibitem{Cockburn2009}
A.~Cockburn, A.~Karlson, and B.~B. Bederson.
\newblock A review of overview+detail, zooming, and focus+context interfaces.
\newblock {\em ACM Computing Surveys}, 41(1):1--31, 2008.

\bibitem{cordeil2019iatk}
M.~Cordeil, A.~Cunningham, B.~Bach, C.~Hurter, B.~H. Thomas, K.~Marriott, and T.~Dwyer.
\newblock Iatk: An immersive analytics toolkit.
\newblock In {\em 2019 IEEE conference on virtual reality and 3D user interfaces (VR)}, pp. 200--209. IEEE, 2019.

\bibitem{Elmqvist2008}
N.~Elmqvist and P.~Tsigas.
\newblock A taxonomy of {3D} occlusion management for visualization.
\newblock {\em IEEE Transactions on Visualization and Computer Graphics}, 14(5):1095--1109, 2008.

\bibitem{Fan2025}
S.~Fan, R.~Liu, W.~Wang, and Y.~Yang.
\newblock Scene map-based prompt tuning for navigation instruction generation.
\newblock In {\em Proceedings of the CVPR}, pp. 6898--6908, 2025.

\bibitem{Garth2009}
C.~Garth, G.-S. Li, X.~Tricoche, C.~D. Hansen, and H.~Hagen.
\newblock Visualization of coherent structures in transient 2d flows.
\newblock In {\em Topology-Based Methods in Visualization II}, pp. 1--13. Springer, 2009.

\bibitem{gruenefeld2019comparing}
U.~Gruenefeld, I.~Koethe, D.~Lange, S.~Wei{\ss}, and W.~Heuten.
\newblock Comparing techniques for visualizing moving out-of-view objects in head-mounted virtual reality.
\newblock In {\em 2019 IEEE conference on virtual reality and 3D user interfaces (VR)}, pp. 742--746. IEEE, 2019.

\bibitem{hart1988development}
S.~G. Hart and L.~E. Staveland.
\newblock Development of nasa-tlx (task load index): Results of empirical and theoretical research.
\newblock In {\em Advances in psychology}, vol.~52, pp. 139--183. Elsevier, 1988.

\bibitem{OpenAI2024GPT4o}
A.~Hurst, A.~Lerer, A.~P. Goucher, A.~Perelman, A.~Ramesh, A.~Clark, et~al.
\newblock {GPT-4o System Card}.
\newblock {\em arXiv preprint arXiv:2410.21276}, 2024.

\bibitem{issartel2016tangible}
P.~Issartel, L.~Besan{\c{c}}on, T.~Isenberg, and M.~Ammi.
\newblock A tangible volume for portable 3d interaction.
\newblock In {\em 2016 IEEE International Symposium on Mixed and Augmented Reality (ISMAR-Adjunct)}, pp. 215--220. IEEE, 2016.

\bibitem{Julier2000}
S.~Julier et~al.
\newblock Information filtering for mobile augmented reality.
\newblock In {\em Proceedings of the ISAR}, 2000.

\bibitem{keshavarz2011validating}
B.~Keshavarz and H.~Hecht.
\newblock Validating an efficient method to quantify motion sickness.
\newblock {\em Human factors}, 53(4):415--426, 2011.

\bibitem{OpenSciVis}
P.~Klacansky.
\newblock Open scivis datasets, 2017.
\newblock Accessed: 2026-03.

\bibitem{Kraus2022}
M.~Kraus et~al.
\newblock The value of immersive visualization.
\newblock {\em IEEE Computer Graphics and Applications}, 42(4):18--28, 2022.

\bibitem{Langner2021}
R.~Langner et~al.
\newblock {MARVIS}: Combining mobile devices and augmented reality for visual data analysis.
\newblock {\em IEEE Transactions on Visualization and Computer Graphics}, 27(2):833--843, 2021.

\bibitem{Laramee2004}
R.~S. Laramee et~al.
\newblock The state of the art in flow visualization: Dense and texture-based techniques.
\newblock {\em Computer Graphics Forum}, 23(2):203--221, 2004.

\bibitem{Lazar2017research}
J.~Lazar, J.~H. Feng, and H.~Hochheiser.
\newblock {\em Research methods in human-computer interaction}.
\newblock Morgan Kaufmann, 2017.

\bibitem{Li2023ImmerView}
J.~Li, L.~Zhao, H.-N. Liang, and L.~Yu.
\newblock Immerview: Adaptive multi-view layout for immersive situated visualizations.
\newblock In {\em 2023 IEEE International Symposium on Mixed and Augmented Reality Adjunct (ISMAR-Adjunct)}, pp. 108--112. IEEE, 2023.

\bibitem{Li2025}
N.~Li, Z.~He, L.~Shen, T.~Wu, B.~Gao, Y.~Zhang, L.~Zhang, F.~Tian, T.~Luo, T.~Han, and Q.~Wang.
\newblock A dual-stick controller for enhancing raycasting interactions with virtual objects.
\newblock In {\em 2025 IEEE Conference Virtual Reality and 3D User Interfaces (VR)}. IEEE, 2025.

\bibitem{Li2023}
Z.~Li et~al.
\newblock {NeRF-Editing}: Geometry editing of neural radiance fields.
\newblock In {\em Proceedings of the CVPR}, 2023.

\bibitem{liu2025dgdiff}
Y.~Liu, X.~Zhang, Q.~Geng, and Z.~Zhou.
\newblock Dgdiff: Immersive 3d indoor scene synthesis via dialog-graph conditioned diffusion.
\newblock In {\em 2025 IEEE International Symposium on Mixed and Augmented Reality (ISMAR)}, pp. 444--453. IEEE, 2025.

\bibitem{Meyer2008}
B.~C. Meyer, T.~Werncke, W.~Hopfenmüller, H.~J. Raatschen, K.~J. Wolf, and T.~Albrecht.
\newblock Dual energy ct of peripheral arteries: effect of automatic bone and plaque removal on image quality and grading of stenoses.
\newblock {\em European Journal of Radiology}, 68(3):414--422, 2008.

\bibitem{Mota2018}
R.~C. Mota, A.~Rocha, J.~D. Silva, U.~Alim, and E.~Sharlin.
\newblock 3de interactive lenses for visualization in virtual environments.
\newblock In {\em 2018 IEEE Scientific Visualization Conference (SciVis)}, pp. 21--25. IEEE, 2018.

\bibitem{Oeltze2006}
S.~Oeltze et~al.
\newblock Integrated visualization of morphologic and perfusion data for the analysis of coronary artery disease.
\newblock In {\em EuroVis}, 2006.

\bibitem{o2024immersive}
S.~O’Keeffe, B.~H. Thomas, J.~O’Hehir, J.~Rombouts, M.~Balasso, and A.~Cunningham.
\newblock Immersive focus+ context techniques to assist in interpretation of high density forest point clouds.
\newblock In {\em 2024 IEEE International Symposium on Mixed and Augmented Reality (ISMAR)}, pp. 961--970. IEEE, 2024.

\bibitem{parasuraman1997humans}
R.~Parasuraman and V.~Riley.
\newblock Humans and automation: Use, misuse, disuse, abuse.
\newblock {\em Human factors}, 39(2):230--253, 1997.

\bibitem{Preim2013}
B.~Preim and C.~P. Botha.
\newblock {\em Visual Computing for Medicine: Theory, Algorithms, and Applications}.
\newblock Morgan Kaufmann, 2013.

\bibitem{Qiao2025}
Y.~Qiao et~al.
\newblock {Open-nav}: Exploring zero-shot vision-and-language navigation in continuous environment with open-source llms.
\newblock In {\em ICRA}, pp. 6710--6717, 2025.

\bibitem{Rocha2018}
A.~Rocha, J.~D. Silva, U.~R. Alim, S.~Carpendale, and M.~C. Sousa.
\newblock Decal-lenses: Interactive lenses on surfaces for multivariate visualization.
\newblock {\em IEEE Transactions on Visualization and Computer Graphics}, 25(8):2568--2582, 2018.

\bibitem{sarkar1994graphical}
M.~Sarkar and M.~H. Brown.
\newblock Graphical fisheye views.
\newblock {\em Communications of the ACM}, 37(12):73--83, 1994.

\bibitem{Song2023}
C.~H. Song et~al.
\newblock {LLM-planner}: Few-shot grounded planning for embodied agents with large language models.
\newblock In {\em Proceedings of the IEEE/CVF International Conference on Computer Vision}, pp. 2998--3009, 2023.

\bibitem{Song2025}
X.~Song et~al.
\newblock Towards long-horizon vision-language navigation: Platform, benchmark and method.
\newblock In {\em Proceedings of the CVPR}, pp. 12078--12088, 2025.

\bibitem{Stoakley1995}
R.~Stoakley, M.~J. Conway, and R.~Pausch.
\newblock Virtual reality on a {WIM}: Interactive worlds in miniature.
\newblock In {\em Proceedings of the CHI Conference}, pp. 265--272, 1995.

\bibitem{taibo2024immersive}
J.~Taibo and J.~A. Iglesias-Guitian.
\newblock Immersive 3d medical visualization in virtual reality using stereoscopic volumetric path tracing.
\newblock In {\em 2024 IEEE Conference Virtual Reality and 3D User Interfaces (VR)}, pp. 1044--1053. IEEE, 2024.

\bibitem{tominski2017interactive}
C.~Tominski, S.~Gladisch, U.~Kister, R.~Dachselt, and H.~Schumann.
\newblock Interactive lenses for visualization: An extended survey.
\newblock In {\em Computer Graphics Forum}, vol.~36, pp. 173--200. Wiley Online Library, 2017.

\bibitem{Wang2025TeamPortal}
X.~Wang, L.~Shen, L.~Chen, M.~Fan, and L.-H. Lee.
\newblock Teamportal: Exploring virtual reality collaboration through shared and manipulating parallel views.
\newblock {\em IEEE Transactions on Visualization and Computer Graphics}, 31(5):3314--3324, 2025.

\bibitem{Weiskopf2003}
N.~Weiskopf, R.~Veit, M.~Erb, K.~Mathiak, W.~Grodd, R.~Goebel, and N.~Birbaumer.
\newblock Physiological self-regulation of regional brain activity using real-time functional magnetic resonance imaging (fmri): methodology and exemplary data.
\newblock {\em NeuroImage}, 19(3):577--586, 2003.

\bibitem{Xu2025}
W.~Xu, Y.~Wei, X.~Hu, W.~Stuerzlinger, Y.~Wang, and H.-N. Liang.
\newblock Predicting ray pointer landing poses in vr using multimodal lstm-based neural networks.
\newblock In {\em 2025 IEEE Conference Virtual Reality and 3D User Interfaces (VR)}, pp. 93--103. IEEE, 2025. \href{https://doi.org/10.1109/VR59515.2025.00034}
{doi: {{%
10\hspace{.1pt}\discretionary{.}{%
}{.}\hspace{.4pt}1109\discretionary{/}{%
}{/}VR59515\hspace{.1pt}\discretionary{.}{%
}{.}\hspace{.4pt}2025\hspace{.1pt}\discretionary{.}{%
}{.}\hspace{.4pt}00034}}}


\bibitem{Yang2024Prompt}
S.~Yang, Y.~H. Tsui, X.~Wang, A.~Alhilal, R.~H. Mogavi, X.~Wang, and P.~Hui.
\newblock From prompt to metaverse: User perceptions of personalized spaces crafted by generative ai.
\newblock In {\em Companion Publication of the 2024 Conference on Computer-Supported Cooperative Work and Social Computing}, pp. 497--504, 2024.

\bibitem{Yu2012}
L.~Yu, K.~Efstathiou, P.~Isenberg, and T.~Isenberg.
\newblock Efficient structure-aware selection techniques for {3D} point cloud visualizations with {2DOF} input.
\newblock {\em IEEE Transactions on Visualization and Computer Graphics}, 18(12):2245--2254, 2012.

\bibitem{yu2010fi3d}
L.~Yu, P.~Svetachov, P.~Isenberg, M.~H. Everts, and T.~Isenberg.
\newblock Fi3d: Direct-touch interaction for the exploration of 3d scientific visualization spaces.
\newblock {\em IEEE transactions on visualization and computer graphics}, 16(6):1613--1622, 2010.

\bibitem{Yuan2023MEinVR}
Z.~Yuan, S.~He, Y.~Liu, and L.~Yu.
\newblock Meinvr: Multimodal interaction techniques in immersive exploration.
\newblock {\em Visual Informatics}, 7(3):37--48, 2023.

\bibitem{zhang2023complete}
C.~Zhang, C.~Zhang, S.~Zheng, Y.~Qiao, C.~Li, M.~Zhang, S.~K. Dam, C.~M. Thwal, Y.~L. Tun, L.~L. Huy, et~al.
\newblock A complete survey on generative ai (aigc): Is chatgpt from gpt-4 to gpt-5 all you need?
\newblock {\em arXiv preprint arXiv:2303.11717}, 2023.

\bibitem{Zhang2020Hand}
J.~Zhang, M.~Huang, L.~Zhao, R.~Yang, H.-N. Liang, J.~Han, and W.~Sun.
\newblock Influence of hand representation design on presence and embodiment in virtual environment.
\newblock In {\em 2020 13th International Symposium on Computational Intelligence and Design (ISCID)}, pp. 364--367. IEEE, 2020.

\bibitem{zhang2023survey}
Y.~Zhang, Z.~Wang, J.~Zhang, G.~Shan, and D.~Tian.
\newblock A survey of immersive visualization: Focus on perception and interaction.
\newblock {\em Visual Informatics}, 7(4):22--35, 2023.

\bibitem{Zhao2022LWiM}
L.~Zhao, N.~Cao, S.~He, H.~N. Liang, and L.~Yu.
\newblock L-wim: Collaborative exploration in immersive environments.
\newblock {\em 2022 IEEE International Symposium on Mixed and Augmented Reality Adjunct (ISMAR-Adjunct)}, pp. 118--123, 2022.

\bibitem{Zhao2024}
L.~Zhao et~al.
\newblock {MeTACAST}: Target- and context-aware spatial selection in {VR}.
\newblock {\em IEEE Transactions on Visualization and Computer Graphics}, 30(1):480--494, 2024.

\bibitem{Zhao2024SpatialTouch}
L.~Zhao, T.~Isenberg, F.~Xie, H.-N. Liang, and L.~Yu.
\newblock Spatialtouch: exploring spatial data visualizations in cross-reality.
\newblock {\em IEEE Transactions on Visualization and Computer Graphics}, 31(1):897--907, 2024.

\bibitem{Zhao2026ScaleFree}
L.~Zhao, F.~Xie, T.~Isenberg, H.-N. Liang, and L.~Yu.
\newblock Scalefree: Dynamic kde for multiscale point cloud exploration in vr.
\newblock {\em arXiv preprint arXiv:2601.20758}, 2026.

\bibitem{Zhou2024}
G.~Zhou, Y.~Hong, and Q.~Wu.
\newblock {NavGPT}: Explicit reasoning in vision-and-language navigation with large language models.
\newblock In {\em Proceedings of the AAAI Conference on Artificial Intelligence}, vol.~38, pp. 7641--7649, 2024.

\bibitem{zimmermann2025multi}
E.~Zimmermann and S.~Bruckner.
\newblock Multi-focus probes for context-preserving network exploration and interaction in immersive analytics.
\newblock In {\em 2025 IEEE Visualization and Visual Analytics (VIS)}, pp. 346--350. IEEE, 2025.

\bibitem{zipf2016human}
G.~K. Zipf.
\newblock {\em Human behavior and the principle of least effort: An introduction to human ecology}.
\newblock Ravenio books, 2016.

\end{thebibliography}
\end{document}